\documentclass[conference]{IEEEtran}
\IEEEoverridecommandlockouts
\usepackage{cite}
\usepackage{amsmath,amssymb,amsfonts}
\usepackage{algorithmic}
\usepackage{graphicx}
\usepackage{textcomp}
\usepackage{xcolor}

\usepackage{caption}
\usepackage{tabularx}
\usepackage{subcaption}
\usepackage{threeparttable}

\def\BibTeX{{\rm B\kern-.05em{\sc i\kern-.025em b}\kern-.08em
    T\kern-.1667em\lower.7ex\hbox{E}\kern-.125emX}}
\begin{document}

\title{Leveraging Label Potential for Enhanced Multimodal Emotion Recognition
}

\author{
\IEEEauthorblockA{Xuechun Shao$^1$, Yinfeng Yu$^1$$^{\dagger}, $Liejun Wang$^1$$^{\dagger}$\thanks{$^{\dagger}$Both Yinfeng Yu and Liejun Wang are corresponding authors.} }
\\
\IEEEauthorblockA{$^1$Xinjiang Multimodal Intelligent Processing and Information Security Engineering Technology Research Center, \\ School of Computer Science and Technology, Xinjiang University, China}
\\
\IEEEauthorblockA{E-mail: yuyinfeng@xju.edu.cn, wljxju@xju.edu.cn}
}


\maketitle

\begin{abstract}


Multimodal emotion recognition (MER) seeks to integrate various modalities to predict emotional states accurately. However, most current research focuses solely on the fusion of audio and text features, overlooking the valuable information in emotion labels. This oversight could potentially hinder the performance of existing methods, as emotion labels harbor rich, insightful information that could significantly aid MER. We introduce a novel model called Label Signal-Guided Multimodal Emotion Recognition (LSGMER) to overcome this limitation. This model aims to fully harness the power of emotion label information to boost the classification accuracy and stability of MER. Specifically, LSGMER employs a Label Signal Enhancement module that optimizes the representation of modality features by interacting with audio and text features through label embeddings, enabling it to capture the nuances of emotions precisely. Furthermore, we propose a Joint Objective Optimization(JOO) approach to enhance classification accuracy by introducing the Attribution-Prediction Consistency Constraint (APC), which strengthens the alignment between fused features and emotion categories. Extensive experiments conducted on the IEMOCAP and MELD datasets have demonstrated the effectiveness of our proposed LSGMER model.
\end{abstract}


\section{Introduction}
Emotion is an essential aspect of human life that profoundly influences our thoughts, behaviors, and decision-making. Emotion recognition technology is widely used in chatbots\cite{zhou2020design}, social media analytics\cite{gaind2019emotion}, intelligent customer service\cite{li2019acoustic}, and mental health monitoring\cite{ghosh2019emokey}, etc. It has become a significant research topic in the field of human-computer interaction and plays an important role in enhancing user experience, optimizing the interaction process, and assisting in emotion analysis.


Emotion is inherently multimodal, and a single modality is insufficient to fully capture its complexity. This challenge exists in other domains as well: for example, navigation requires the fusion of visual and auditory modalities to enhance environmental perception\cite{yu2022pay}\cite{yu2023echo}. In emotion recognition, effective integration of multimodal information such as speech and text is essential to improve recognition performance\cite{zhao2023swrr}.

\begin{figure}[t]
    \centering
    \includegraphics[width=0.5\textwidth]{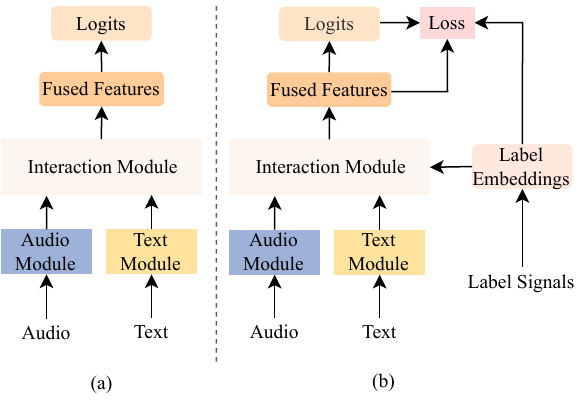}
    \caption{A sketched comparison between the previous main-stream method (a) and the proposed LSGMER (b).}
    \label{fig:difference}
\end{figure}

Currently, most emotion recognition methods use interactions based on attention mechanisms\cite{vaswani2017attention} to fuse multimodal information. For instance, MFCN\cite{MCFN} employs a cross-modal attention mechanism to directly integrate audio and text features, while MSMSER\cite{MSMSER} first computes self-attention for audio and text features separately, then applies an additional attention mechanism again to extract emotional information using learnable query vectors from the concatenated modal features. Although these methods effectively integrate information from different modalities, they still have some limitations. Existing methods mainly focus on the fusion of audio and text features, as illustrated in Fig.~\ref{fig:difference}(a). They overlook the fact that emotion labels not only indicate emotion categories but also contain multidimensional features that can assist models in better capturing complex emotional characteristics. Therefore, how to fully utilize emotion labels is the core issue of this research.

We propose a \textbf{L}abel \textbf{S}ignal-\textbf{G}uided \textbf{M}ultimodal \textbf{E}motion \textbf{R}ecognition (LSGMER) to address the limitations of existing methods in utilizing emotion labels and aligning multimodal features. The model introduces label signals through two key components: First, we design a Label Signal Enhancement Module that explicitly combines label signals with audio and text features, enabling the model to capture emotion features shared across different modalities. At the same time, we apply the moving average(MA) method to further smooth the label embedding updates, which helps the model to better capture fine-grained emotion knowledge. Second, we propose an innovative joint objective optimization(JOO) method that combines the Attribution-Prediction Consistency Constraint (APC) with cross-entropy loss to guide the training process. This method forces the fused features to align with the label embeddings at the entity level, enhancing both the accuracy and the consistency of feature expression. As shown in Fig.~\ref{fig:difference}(b), LSGMER models the multimodal interaction between audio, text, and emotion labels while using label embeddings as anchors to effectively guide the fusion and optimization of emotion features. This design not only strengthens the guiding role of label signals but also significantly improves the performance of multimodal emotion recognition.

The main contributions of this paper are as follows:
\begin{itemize}
\item We propose the LSGMER model, which provides additional emotion category information to the model by explicitly and implicitly introducing emotion label information into the model, thus helping the model to capture emotion features more accurately and optimize emotion feature extraction.
\item We propose two key components: Label Signal Enhancement Module with MA method, and Joint Objective Optimization. These components maximize the guiding effect of label signals and improve the effectiveness of multimodal feature fusion.
\item Experimental results demonstrate that LSGMER outperforms current state-of-the-art baseline methods on the IEMOCAP and MELD datasets, highlighting the superiority of our approach in MER tasks.
\end{itemize}

\section{Related Work}
The attention mechanism has become one of the most widely used approaches in multimodal emotion recognition due to its advantages in information capture and cross-modal association modeling. Many studies propose innovations based on the attention mechanism, achieving significant progress. For example, \cite{zhao2023knowledge} proposes a Bayesian attention mechanism that improves emotion recognition accuracy by incorporating emotion-related external knowledge, helping the model focus more effectively on emotion-related features.  \cite{zhao2023swrr} introduces a sliding-window attention mechanism that limits the scope of attention to reduce redundant computation and interference from extraneous information, enabling the model to dynamically fuse text and audio modalities within the maximum effective feature perception. \cite{ghosh2022mmer} introduces a novel multimodal neural network architecture that captures fine-grained emotion features by combining audio and text information while enhancing the model's expressive capability through multitask learning. \cite{ma2023transformer} proposes a transformer-based model combined with self-distillation, which effectively captures intra-modal and inter-modal interactions and dynamically fuses multimodal information to enhance emotion recognition performance. \cite{shou2024low} introduces a low-rank matching attention method, which aims to solve the issue of intra-modal and inter-modal affective interactions, as well as the problem of excessive computational complexity in existing methods.

Despite the progress made by existing approaches in cross-modal interaction and emotion recognition, most of the studies ignore the guiding role of emotion labels in the process of feature alignment and feature fusion, resulting in models that fail to make full use of the emotion information during feature extraction and information integration, thus limiting the overall performance improvement.

Recently, some studies have begun exploring the application of emotion label information in multimodal emotion recognition. For example, \cite{pan2024gemo} uses a pre-trained text model to extract features from label text, and optimizes the distribution of audio features and label text features through contrastive learning. This approach brings audio features with the same emotion label closer together, while further distancing audio features with different emotion labels. However, directly using label text as entity-level anchor information provides a highly simplified description of emotion, which fails to capture more complex emotional details or subtle emotional changes.

\begin{figure*}[t]
    \centering
    \includegraphics[width=\textwidth]{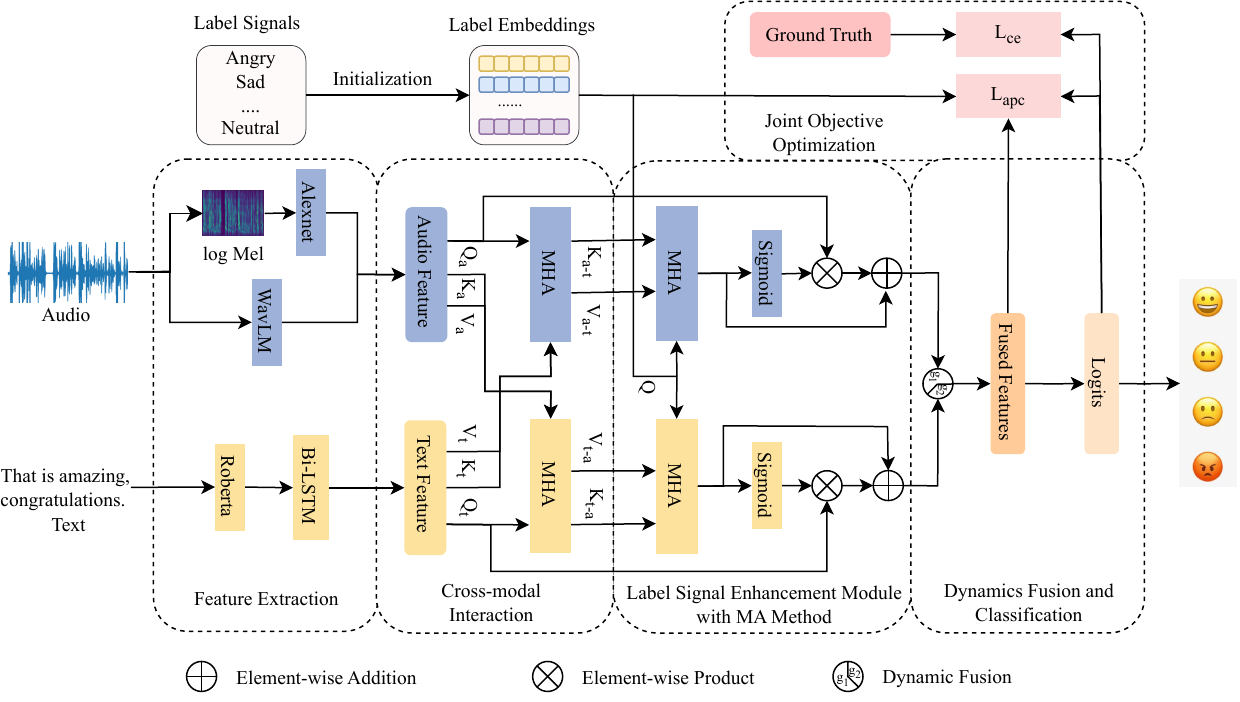}
    \caption{The overall architecture of the LSGMER. MHA refers to Multi-Head Attention, where \( g_{1} \) and \( g_{2} \) represent the learning weights for the audio and text modalities, respectively.}
    \label{fig:LSGMER}
\end{figure*}

\section{Task Definition}
We formulate MER as a classification task. Given a text and its corresponding audio, we extract the corresponding text feature sequence $\textbf{T} = \{(t_i)\}_{i=1}^{n}$ from the text and the corresponding audio feature sequence $\textbf{A} = \{(a_i)\}_{i=1}^{m}$ from the audio, where $n$ is the length of the text sequence and $m$ is the length of the audio sequence. Our goal is to predict emotion categories using both modalities simultaneously.

\section{Methodology}
The overall architecture of the LSGMER model we proposed is shown in Fig.~\ref{fig:LSGMER}. It consists of five main components: Feature Extraction, Cross-modal Interaction, Label Signal Enhancement Module with MA Method, Feature Fusion and Classification, and Joint Objective Optimization.

\subsection{Feature Extraction}
Building on previous work\cite{kim2023focus}, we adopt RoBERTa\cite{liu2019roberta} as the text feature extractor and use the output of its final layer as the text features. These extracted features are then passed through a Bi-LSTM layer to further capture the contextual information in the sequence.
\begin{equation}
 \textbf{H}_{t0} = \mathrm{BiLSTM}(\mathrm{RoBERTa}(\textbf{T})). \label{eq1}
\end{equation}

Considering that audio features contain more information compared to text features\cite{MMRBN}, relying solely on pre-trained models for audio feature extraction may not fully capture the subtle variations in the audio signal. Inspired by the multi-level audio feature extraction method proposed in\cite{zou2022speech}, we combine spectral features with a pre-trained audio model to capture the rich information in the audio signal from multiple levels. To maintain consistency, we use AlexNet to process the spectrogram. At the same time, we use WavLM\cite{chen2022wavlm} to extract high-level features of audio. WavLM focuses more on contextual modeling, shows better performance than other pre-trained models in several audio tasks, and has been widely used in emotion recognition tasks\cite{zhao2023swrr}\cite{MFDR}. Finally, the extracted audio features are summed to create a richer audio feature representation.
\begin{equation}
\textbf{H}_{a0} = \mathrm{WavLM}(\textbf{A}) + \mathrm{AlexNet}(\mathrm{MelSpec}(\textbf{A})). \label{eq2}
\end{equation}


\subsection{Cross-modal Interaction}

The simple sharing of the hidden state is insufficient to achieve effective information transfer between the two modalities. Therefore, we propose a cross-modal attention module that enables the model to capture the relationships between text and audio from multiple perspectives.

Specifically, we use the text features \( \textbf{H}_{t0} \) as the query, and the audio features \( \textbf{H}_{a0} \) as both the key and value to perform cross-modal attention. This process produces text features enriched with audio information:

\begin{equation}
\textbf{H}_{t1} = \mathrm{Attention}(\textbf{H}_{t0}, \textbf{H}_{a0}, \textbf{H}_{a0}). \label{eq3}
\end{equation}

Similarly, we obtain audio features enriched with text information:

\begin{equation}
\textbf{H}_{a1} = \mathrm{Attention}(\textbf{H}_{a0}, \textbf{H}_{t0}, \textbf{H}_{t0}). \label{eq4}
\end{equation}

\subsection{Label Signal Enhancement Module with MA Method}
Although we perform preliminary fusion and alignment of audio and text features in the cross-modal interaction module, the mechanism still has limitations in coping with feature inconsistency and precise alignment of emotion information, which is insufficient to effectively guide the model to focus on the core features of the emotion. Inspired by \cite{jiang2023empathy}, we construct the LSMA module. In this process, label embeddings are used as additional emotion category information to establish associations with audio and text features, respectively, allowing the model to concentrate on key features related to emotion.

To begin, we define a series of label embeddings \(\textbf{L}_{s} \), each corresponding to a specific emotion category. These label embeddings are then used to compute attention with audio features \( \textbf{H}_{a1} \) and text features \( \textbf{H}_{t1} \) respectively. The label embeddings serve as the query, while the audio and text features are used as the key and value. The resulting enhanced audio and text features can be represented as:

\begin{equation}
\textbf{H}_{a2} = \mathrm{Attention}(\textbf{L}_{s}, \textbf{H}_{a1}, \textbf{H}_{a1}),\label{eq5}
\end{equation}
\begin{equation}
\textbf{H}_{t2} = \mathrm{Attention}(\textbf{L}_{s}, \textbf{H}_{t1}, \textbf{H}_{t1}).\label{eq6}
\end{equation}

Next, we perform a weighted fusion of the obtained features with the original features to ensure that the crucial original signals are preserved during the label signal enhancement process. Taking audio features as an example, we apply the sigmoid function to \( \textbf{H}_{a2} \) to obtain the weight vector \( w_a = \sigma(\textbf{H}_{a2}) \). This weight vector is then element-wise multiplied with \( \textbf{H}_{a0} \), and the weighted original audio features are added to \( \textbf{H}_{a2} \) to obtain the final audio features \( \textbf{H}_{a} \):

\begin{equation}
\textbf{H}_a = w_a \odot \textbf{H}_{a0} + \textbf{H}_{a2}. \label{eq7}
\end{equation}

Similarly, the final text features \(\textbf{H}_{t} \) can be expressed as:
\begin{equation}
\textbf{H}_t = w_t \odot \textbf{H}_{t0} + \textbf{H}_{t2}. \label{eq8}
\end{equation}

Since the label embeddings are learnable, they are adapted to the data of the current batch and serve as inputs for the next batch. Therefore the difference in label embeddings at the beginning of each epoch can be substantial, requiring the model to constantly adapt to the new label embeddings, leading to instability in the training process. To alleviate this concern, we employ the MA method to compute the label embeddings.

Specifically, we update the label embeddings using the following formula:
\begin{equation}
\textbf{L}_{t} = \alpha \textbf{L}_{t-1} + (1 - \alpha) \textbf{L}_{t-1}^{'},\label{eq9}
\end{equation}
where \(\alpha\) is a hyperparameter that controls the update rate, \(\textbf{L}_{t-1}\) are the label embeddings at the beginning of epoch \(t-1\), and \(\textbf{L}_{t-1}^{'}\) is the updated label embeddings after epoch \(t-1\). When \(t = 1\), the label embeddings are randomly initialized. 

This update method ensures that historical information effectively guides the current learning process while preventing excessive fluctuations in label embeddings during training.

\subsection{Dynamics Fusion and Classification}
We perform the fusion of audio and text features using an expert gate\cite{aljundi2017expert} to dynamically adjust their contributions to emotion recognition.

The features \(\textbf{H}_{a}\) and \(\textbf{H}_{t}\) are concatenated into a combined feature vector \(\textbf{H}_{\text{concat}}\), and the dynamic weights are computed through a fully connected layer, with the output values normalized by the softmax function. The calculation is as follows:
\begin{equation}
g_{a}, g_{t} = \mathrm{softmax}(w \cdot \textbf{H}_{\text{concat}} + b),\label{eq10}
\end{equation}
where \(g_a\) and \(g_t\) are the dynamic weights of audio features and text features, respectively.

The audio and text features are then weighted according to these weights to obtain the final fused feature representation:
\begin{equation}
\textbf{H}_{\text{fused}} = g_a \odot \textbf{H}_a + g_t \odot \textbf{H}_t,\label{eq11}
\end{equation}
where \(\odot\) denotes element-wise multiplication.

The fused feature \(\textbf{H}_{\text{fused}}\) is subsequently passed through a Multi-Layer Perceptron to compute the emotion category scores for the samples. Next, the scores of each sample are normalized using the softmax function to get the predicted probability for each emotion category:
\begin{equation}
\textit{logits} = \mathrm{MLP}(\textbf{H}_{\text{fused}}),\label{eq12}
\end{equation}
\begin{equation}
p = \mathrm{softmax}(\textit{logits}).\label{eq13}
\end{equation}

\subsection{Joint Objective Optimization}
\begin{figure}[t]
\centering
\includegraphics[width=0.5\textwidth]{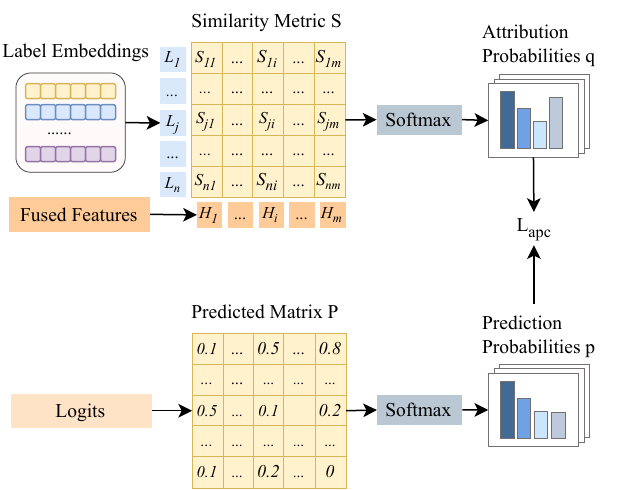}
\caption{The Attribution-Prediction Consistency Constraint.}
\label{fig:apc}
\end{figure}

Feature fusion can effectively compensate for the limitations of a single modality, but ensuring that the fused features accurately represent emotion category information remains a challenge due to the lack of effective supervisory signals. To address this issue, we propose a JOO that combines the Attribution-Predictive Consistency  Constraint (APC)\cite{huang2023pram} and the cross-entropy loss to guide the model training. The schematic of APC loss is shown in Fig.~\ref{fig:apc}. The APC loss minimizes the difference between the similarity distribution of the fused features and label embeddings and the model predictive distribution. It ensures that when the model performs multimodal feature fusion, the learned features can better express emotions, thus improving the classification accuracy.

Specifically, we compute the cosine similarity between fused features and label embeddings:

\begin{equation}
s(\textbf{H}_i, \textbf{L}_j) = \cos(\textbf{H}_i, \textbf{L}_j) = \frac{\textbf{H}_i \cdot \textbf{L}_j}{\|\textbf{H}_i\| \cdot \|\textbf{L}_j\|}.\label{eq14}
\end{equation}

Next, the similarity distribution \(\textbf{q}_i=\{ \textbf{q}(i,j) \}_{j=1}^N\) is produced by applying the softmax function on the cosine similarity score distribution:
\begin{equation}
\textbf{q}(i, j) = \frac{\exp(s(\textbf{H}_i, \textbf{L}_j))}{\sum_{j=1}^N \exp(s(\textbf{H}_i, \textbf{L}_{j}))}.\label{eq15}
\end{equation}
where \(\textbf{q}(i,j)\) represents the probability of similarity between the \(i\)-th sample and the \(j\)-th emotion label, reflecting the likelihood that the sample belongs to the emotion category.

Finally, the APC improves the consistency between the similarity distribution \( \textbf{q}_i \) and the predictive probability distribution \( \textbf{p}_i \) by minimizing the Jensen-Shannon divergence. The objective function can be expressed as:

\begin{equation}
L_{\text{apc}} = \frac{1}{MN} \sum_{i=0}^{M} \sum_{j=0}^{N} \text{JsDiv}(\textbf{q}_{i,j} \parallel \textbf{p}_{i,j}), \label{eq16}
\end{equation}
\begin{align}
\text{JsDiv}(\textbf{q} \parallel \textbf{p}) &= \frac{1}{2} \sum \textbf{p} \log\left( \frac{2\textbf{p}}{\textbf{p} + \textbf{q}} \right) \nonumber \\
&\quad + \frac{1}{2} \sum \textbf{q} \log\left( \frac{2\textbf{q}}{\textbf{p} + \textbf{q}} \right),\label{eq17}
\end{align}
where \(M\) is the training data size and \(N\) is the number of emotion categories.

Then, we utilize cross-entropy loss to estimate the quality of emotion prediction:
\begin{equation}
L_{\text{ce}} = - \frac{1}{M} \sum_{i=1}^{M} \sum_{j=1}^{N} \textbf{y}_{i,j} \log(\textbf{p}_{i,j}),\label{eq18}
\end{equation}
where \(\textbf{y}_{i,j}\) is the one-hot encoding of the ground truth.


Finally, the APC loss and the cross-entropy loss together constitute the total loss of the LSGMER framework. The total loss is calculated as follows:
\begin{equation}
L(\theta) = \alpha L_{\text{ce}}(\theta) + \beta L_{\text{apc}}(\theta),\label{eq19}
\end{equation}
where \(\alpha\) and \(\beta\) are the balancing weights.

\section{Experiments}

\subsection{Dataset}
We use the IEMOCAP\cite{busso2008iemocap} and MELD\cite{poria2018meld} datasets to evaluate the proposed model.

The IEMOCAP dataset contains approximately 12 hours of data, divided into five sessions, each consisting of two speakers. All the discourses are labeled with one of ten emotions: angry, happy, sad, neutral, frustrated, excited, fearful, surprised, disgusted, and other. Following previous research, we perform a categorization task on 5531 discourses, focusing on four emotional categories: happy (combined with excited), angry, sad, and neutral. We conduct our experiments in the 5-fold leave-one-session-out strategy. One session at a time is selected as the test set and the remaining four serve as the training set.

The MELD dataset is a multi-party dataset created from the Friends TV series. The dataset contains about 13,000 discourses. Each is labeled with one of the following seven emotions: anger, disgust, sad, joy, neutral, surprise and fear. The dataset is divided into train, valid, and test sets. In this experiment, we use only the train and test sets.

\subsection{Implementation Details}

We extract 768-dimensional text embeddings and 1024-dimensional audio embeddings using the pre-trained models RoBERTa and WavLM, and employ AdamW as the optimizer. On the IEMOCAP dataset, the learning rate is set to 5e-4, and $\alpha$ and $\beta$ are set to 1 and 0.05, respectively. On the MELD dataset, the learning rate is set to 5e-4, and $\alpha$ and $\beta$ are set to 1 and 0.1, respectively. In the MA method, the hyperparameter controlling the update rate is set to 0.99. We train the model on these two datasets for 50 epochs respectively. 

We use Weighted Accuracy (WA), Unweighted Accuracy (UA) and Weighted F1-Score (WF1) as evaluation metrics.

\subsection{Baselines}
MER-HAN\cite{MER-HAN} combines three attention mechanisms, local intra-modal attention, cross-modal attention, and global inter-modal attention to effectively learn emotional features.

DWFORMER\cite{DWFORMER} proposes a dynamic local window transformer module and a dynamic global window transformer module to fully utilize local details and global context information.

LSTM-Attn\cite{LSTM-Attn} introduces a novel strategy for feature pooling over time which uses local attention in order to focus on specific regions of a speech signal that are more emotionally salient.


KS-Transformer\cite{KS-Transformer} proposes a sparse transformer model that can focus on the key emotional information from different modalities during cross-modal computation, while also reducing redundant calculations.

MSMSER\cite{MSMSER} designs a set of trainable emotion tokens to retrieve emotion information from concatenated audio and text features. Randomly masking a modality during training forces the model to fully perceive the emotional features in each modality.

MMRBN\cite{MMRBN} proposes that audio is more important than text in emotion recognition. It utilizes cross-modal attention to fuse features between modalities, and then dynamically combines emotional information from each modality.

MFDR\cite{MFDR} proposes sliding adaptive window attention for modeling the acoustic-text fusion stage, using dynamic frame convolution to identify and weaken fine-grained information that is irrelevant to emotional expression.

DST\cite{DST} introduces a deformable speech transformer that adaptively determines the size and position of the attention window to reduce redundant computation.

MSTR\cite{MSTR} utilizes a multi-scale transformer to capture emotional information in speech, effectively capturing the variations of emotion across different temporal scales, thereby enhancing the accuracy of emotion recognition.


MM-DFN\cite{MM-DFN} designs a new graph-based dynamic fusion module that helps to fuse multimodal information and enhance the complementarity between modalities.

MCFN\cite{MCFN} proposes a two-stage fusion network that first learns audio and text features separately and then fuses them through a modality collaborative learning module.

GateM\textsuperscript{2}Former\cite{xu2025gatem} dynamically integrates representations from different layers of a pre-trained model through a gating mechanism, combining different modality experts and effectively enhancing model flexibility.

\begin{table}[b]
\begin{center}
\caption{Model performance comparison on IEMOCAP.}
\label{tab:iemocap_results} 
\renewcommand\arraystretch{1.05}
\setlength{\tabcolsep}{3.65mm}{
\begin{tabular}{lcccc}
\hline
\textbf{Models} & \textbf{Year} & \textbf{Modality} & \textbf{WA\%} & \textbf{UA\%} \\
\hline
MER-HAN & 2023 & A & 55.5 & 57.0 \\
DWFORMER  & 2023 & A & 72.3 & 73.9 \\
\hline
LSTM-Attn & 2017 & T & 63.3 & 63.5 \\
MER-HAN & 2023 & T & 68.6 & 69.5 \\
\hline
KS-Transformer & 2022 & A+T & 74.3 & 75.3 \\
MER-HAN & 2023 & A+T & 73.3 & 74.2 \\
MSMSER & 2023 & A+T & 75.2 & 76.4 \\
MMRBN & 2024 & A+T & 76.0 & 76.6 \\
MFDR & 2024 & A+T & 75.7 & 77.0 \\
GateM\textsuperscript{2}Former & 2024 & A+T & 75.9 & 77.4  \\
\hline
LSGMER (Ours) & 2024 & A+T & \textbf{77.5} & \textbf{77.9} \\
\hline
\end{tabular}}

\begin{tablenotes}                
    \item Note: The baseline results are directly cited from their original papers, and the best results in the table are highlighted in bold. A and T are audio and text modalities, respectively.
\end{tablenotes}
\end{center}
\end{table}
\subsection{Results and Discussion}
Tables~\ref{tab:iemocap_results} and~\ref{tab:meld_results} present the performance of the baseline models and LSGMER on the IEMOCAP and MELD datasets, respectively. On the IEMOCAP dataset, LSGMER outperforms all baseline models, achieving improvements of 2.11\% in WA and 0.65\% in UA compared to GateM\textsuperscript{2}Former. On the MELD dataset, the LSGMER model also demonstrates leading performance, achieving a WF1 score of 62.9\%, which is significantly higher than the other comparative models, with a 1.13\% improvement over MCFN. These results further validate the superiority and wide applicability of LSGMER in multimodal emotion recognition tasks.


We attribute the performance improvement to the guiding role of emotion labels in the model. We introduce emotion label information into the model, and emotion label embeddings provide explicit emotion category anchors during the training process. This approach is different from traditional methods, which rely only on audio and text features for training and ignore the rich information embedded in emotion labels. Guided by the emotion label embeddings, the model is able to capture the emotion features more accurately and make more appropriate decisions when fusing multimodal information. At the same time, the APC loss can ensure that the features learned by the model match the emotion label embeddings. This process can effectively improve the accuracy of emotion recognition and reduce redundant information in multimodal information fusion, ultimately realizing more efficient and accurate emotion recognition.

\begin{table}[t]
\begin{center}
\caption{Model performance comparison on MELD.}
\label{tab:meld_results} 
\renewcommand\arraystretch{1.05}
\setlength{\tabcolsep}{3.65mm}{
\begin{tabular}{lcccc}
\hline
\textbf{Models} & \textbf{Year} & \textbf{Modality} & \textbf{WF1\%}  \\
\hline
DWFORMER  & 2023 & A & 48.5  \\
DST & 2023 & A & 48.8  \\
MSTR & 2023 & A & 46.1  \\
\hline
MM-DFN & 2022 & A+T & 58.3  \\
MCFN & 2023 & A+T & 62.2  \\
GateM\textsuperscript{2}Former & 2024 & A+T & 61.2  \\
\hline
LSGMER (Ours) & 2024 & A+T & \textbf{62.9}  \\
\hline
\end{tabular}}
\begin{tablenotes}                
    \item Note: The baseline results are directly cited from their original papers, and the best results in the table are highlighted in bold. A and T are audio and text modalities, respectively.
\end{tablenotes}
\end{center}
\end{table}

\subsection{Ablation Study}

\begin{table}[t]
\begin{center}
\caption{Ablation study on the proposed model.}
\label{tab:ablation} 
\renewcommand\arraystretch{1.00}
\setlength{\tabcolsep}{3.65mm}{
\begin{threeparttable} 
\begin{tabular}{lccc}
\hline
        & \multicolumn{2}{c}{\textbf{IEMOCAP}} & \textbf{MELD} \\
\hline
\textbf{Models}  & \textbf{WA\%} & \textbf{UA\%} & \textbf{WF1\%} \\
\hline
LSGMER & \textbf{77.5} & \textbf{77.9} & \textbf{62.9}\\
\hline
w/o MA   & 76.7 & 77.4 & 62.1 \\
w/o JOO   & 76.5 & 76.9 & 61.8\\
w/o LSMA \& JOO  & 74.1 & 75.2 & 60.9\\
\hline
\end{tabular}
\begin{tablenotes}                
    \item Note: “w/o” denotes without.
\end{tablenotes}
\end{threeparttable}}
\end{center}
\end{table}

We carry out ablation experiments on the IEMOCAP and MELD datasets. Table~\ref{tab:ablation} lists the results under various ablation settings.

1) w/o MA: The MA method is excluded from the update process for label embeddings. After each training epoch, the current label embeddings are directly initialized as label embeddings for the next epoch.

Removing the MA method degrades model performance, demonstrating that stable updates of label embeddings are crucial for model effectiveness. Without MA, the label embeddings exhibit drastic fluctuations during training cycles. Such instability may prevent the model from effectively capturing consistent emotion representations during training, ultimately compromising prediction accuracy.

2) w/o JOO: The APC loss is removed, and the model is trained using only the cross-entropy loss.

The absence of JOO results in a significant drop in model performance, indicating that JOO provides crucial supervisory signals during training. The APC loss helps the model better understand the structure of emotional categories by optimizing the similarity between label embeddings and features, and effectively aligns features with label embeddings. Without JOO, the model fails to fully exploit the potential of label embeddings, leading to a reduction in emotion recognition capabilities. Therefore, as a core component of JOO, APC loss is instrumental in enhancing the model emotion classification performance.

3) w/o LSMA \& JOO: Both the LSMA module and JOO are removed, meaning the model no longer receives guidance from label signals and relies solely on audio and text features for training.

Removing the LSMA module and JOO shows a significant decrease in model performance, indicating that these two modules play a crucial role in emotion recognition tasks. The LSMA module utilizes label embeddings to align audio and text features, allowing the model to focus on key features related to emotion. Meanwhile, JOO further strengthens the guidance of label signals. Without these two modules, the model can only rely on audio and text features for training, failing to accurately capture emotional information, which negatively impacts the accuracy of emotion classification. This experiment further emphasizes the importance of label signals, as they provide effective guidance for emotion categories and enhance the overall recognition performance of the model.

The results of these ablation experiments demonstrate that each module is essential and works synergistically to enhance the model's ability to recognize emotion.

\subsection{Visualization}
\begin{figure}[t] 
    \centering  
    \begin{minipage}[t]{0.23\textwidth} 
        \centering
        \includegraphics[width=\textwidth]{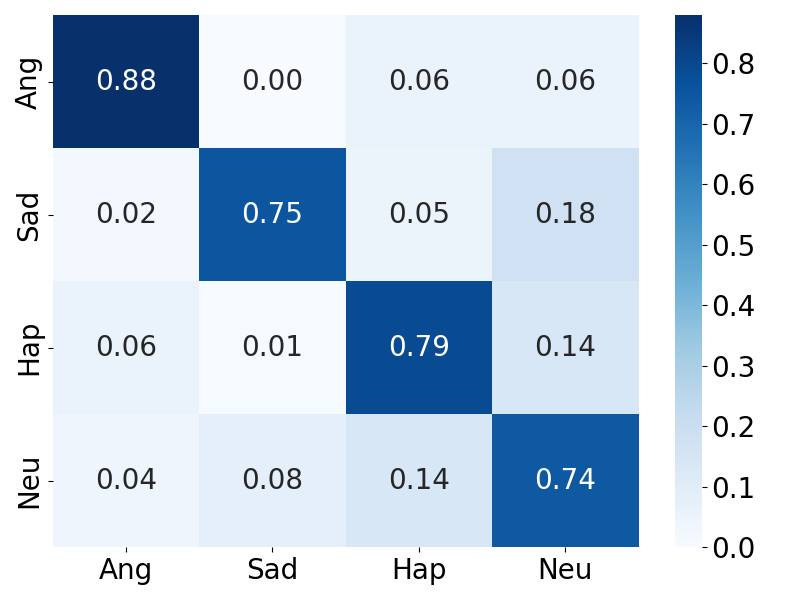} 
        \subcaption{w/o JOO \& LSMA}
        \label{fig:abl3_confusion}
    \end{minipage}%
    \hfill 
    \begin{minipage}[t]{0.23\textwidth} 
        \centering
        \includegraphics[width=\textwidth]{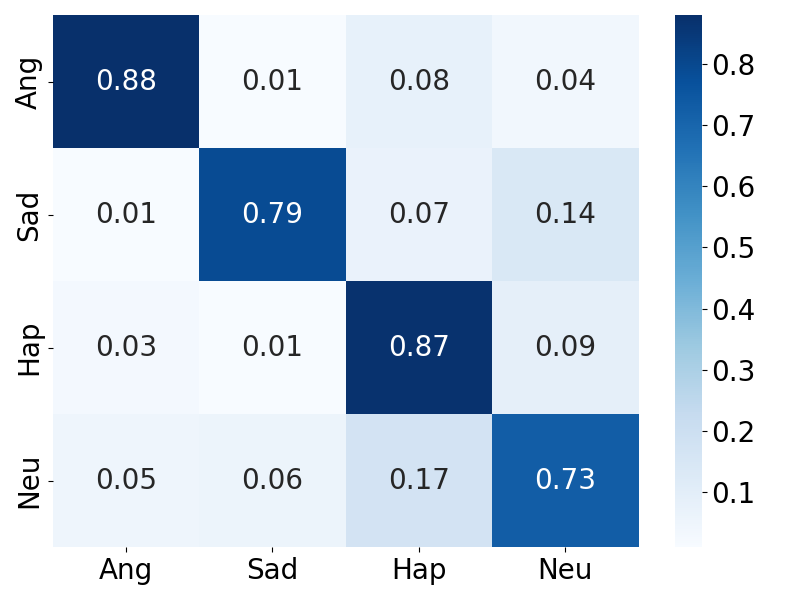} 
        \subcaption{LSGMER}
        \label{fig:LSGMER_confusion}
    \end{minipage}

    \caption{Comparison of normalized confusion matrices on the IEMOCAP dataset.} 
    \label{fig:confusion matrices}

    \vspace{1em} 

     \begin{minipage}[t]{0.24\textwidth} 
        \centering
        \includegraphics[width=\textwidth]{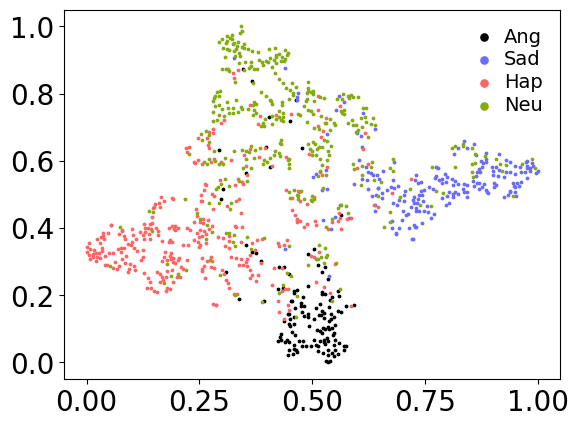} 
        \subcaption{w/o JOO \& LSMA}
        \label{fig:abl3_visualization}
    \end{minipage}%
    \hfill 
    \begin{minipage}[t]{0.24\textwidth} 
        \centering
        \includegraphics[width=\textwidth]{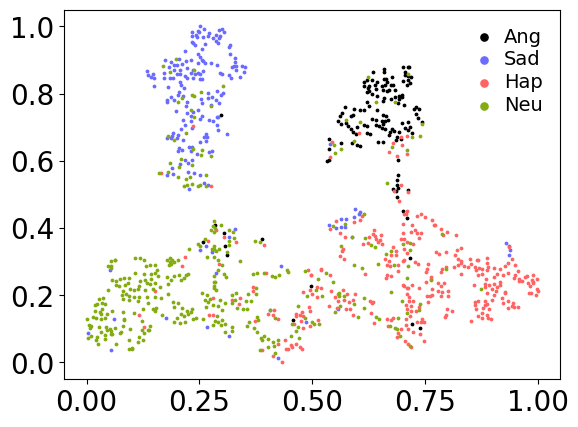} 
        \subcaption{LSGMER}
        \label{fig:LSGMER_visualization}
    \end{minipage}
    
    \vspace{1em} 
    \caption{The t-SNE visualization of feature distribution on the IEMOCAP dataset.} 
    \label{fig:t-SNE visualization}
\end{figure}

Fig.~\ref{fig:confusion matrices} illustrates the normalized confusion matrix comparison of LSGMER on the IEMOCAP dataset with and without label guidance. The results clearly show that the recognition precisions of emotion categories are significantly improved with the label guidance, especially for the categories of sadness and happiness, which are increased by 4\% and 8\%, respectively. These enhancements indicate that label embeddings help the model to distinguish and recognize emotions more clearly by providing additional emotion category information to the model, which reduces the confusion between emotion categories, especially the crossover between neutral, sad and happy. By integrating label embeddings, the model is able to optimize feature representations more effectively across categories during the learning process, thereby boosting the overall recognition accuracy.

We extract the fused feature of each discourse on IEMOCAP from LSGMER with or without label guidance. As can be seen in Fig.~\ref{fig:t-SNE visualization}, when there is no guidance from the label embeddings, the distribution between emotion categories is more chaotic, and there is a large overlap between different emotions, which leads to the model's difficulty in distinguishing these emotion categories. After adding label signals, the distribution of emotion categories in the feature space is more dispersed, the boundaries between emotion categories are more obvious, and the overlapping region is significantly reduced. This demonstrates that the guiding effect of label signals enables the model to better optimize the boundaries between emotion categories and improves the separability of the features.

\section{Conclusion}
In this paper, we introduce a novel label signal-guided MER architecture that innovatively incorporates LSMA and JOO modules. These modules effectively introduce label signals into the model, providing additional emotion category information and utilizing label embeddings to align features across different modalities. This enables the model to process multimodal data, such as audio and text, more efficiently and to achieve more accurate emotion classification. Experimental results on two datasets demonstrate that our proposed model achieves new state-of-the-art performance.

For future work, we will improve the generalization ability of the model and extend its application to real-world scenarios. Drawing on research in the field of audio-visual navigation \cite{SAAVN}, which focuses on improving robustness in dynamic and noisy environments, we would like to employ a similar strategy to enhance the adaptability of emotion recognition systems in complex environments. This will enable us to maintain efficient emotion categorization and decision-making capabilities in the face of various disturbances and environmental noise.

\section{Acknowledgements}

This study was funded by the Excellence Program Project of Tianshan, Xinjiang Uygur Autonomous Region, China (grant number 2022TSYCLJ0036), the Central Government Guides Local Science and Technology Development Fund Projects (grant number ZYYD2022C19), and the National Natural Science Foundation of China (grant numbers 62463029, 62472368 and 62303259).

\bibliographystyle{IEEEtran}
\bibliography{references}

\begin{thebibliography}{10}
\providecommand{\url}[1]{#1}
\csname url@samestyle\endcsname
\providecommand{\newblock}{\relax}
\providecommand{\bibinfo}[2]{#2}
\providecommand{\BIBentrySTDinterwordspacing}{\spaceskip=0pt\relax}
\providecommand{\BIBentryALTinterwordstretchfactor}{4}
\providecommand{\BIBentryALTinterwordspacing}{\spaceskip=\fontdimen2\font plus
\BIBentryALTinterwordstretchfactor\fontdimen3\font minus \fontdimen4\font\relax}
\providecommand{\BIBforeignlanguage}[2]{{%
\expandafter\ifx\csname l@#1\endcsname\relax
\typeout{** WARNING: IEEEtran.bst: No hyphenation pattern has been}%
\typeout{** loaded for the language `#1'. Using the pattern for}%
\typeout{** the default language instead.}%
\else
\language=\csname l@#1\endcsname
\fi
#2}}
\providecommand{\BIBdecl}{\relax}
\BIBdecl

\bibitem{zhou2020design}
L.~Zhou, J.~Gao, D.~Li, and H.-Y. Shum, ``The design and implementation of xiaoice, an empathetic social chatbot,'' \emph{Computational Linguistics}, vol.~46, no.~1, pp. 53--93, 2020.

\bibitem{gaind2019emotion}
B.~Gaind, V.~Syal, and S.~Padgalwar, ``Emotion detection and analysis on social media,'' \emph{arXiv preprint arXiv:1901.08458}, 2019.

\bibitem{li2019acoustic}
B.~Li, D.~Dimitriadis, and A.~Stolcke, ``Acoustic and lexical sentiment analysis for customer service calls,'' in \emph{ICASSP 2019-2019 IEEE International Conference on Acoustics, Speech and Signal Processing (ICASSP)}.\hskip 1em plus 0.5em minus 0.4em\relax IEEE, 2019, pp. 5876--5880.

\bibitem{ghosh2019emokey}
S.~Ghosh, S.~Sahu, N.~Ganguly, B.~Mitra, and P.~De, ``Emokey: An emotion-aware smartphone keyboard for mental health monitoring,'' in \emph{2019 11th international conference on communication systems \& networks (COMSNETS)}.\hskip 1em plus 0.5em minus 0.4em\relax IEEE, 2019, pp. 496--499.

\bibitem{yu2022pay}
Y.~Yu, L.~Cao, F.~Sun, X.~Liu, and L.~Wang, ``Pay self-attention to audio-visual navigation,'' \emph{arXiv preprint arXiv:2210.01353}, 2022.

\bibitem{yu2023echo}
Y.~Yu, L.~Cao, F.~Sun, C.~Yang, H.~Lai, and W.~Huang, ``Echo-enhanced embodied visual navigation,'' \emph{Neural Computation}, vol.~35, no.~5, pp. 958--976, 2023.

\bibitem{zhao2023swrr}
Z.~Zhao, T.~Gao, H.~Wang, and B.~W. Schuller, ``Swrr: Feature map classifier based on sliding window attention and high-response feature reuse for multimodal emotion recognition,'' in \emph{Proc. INTERSPEECH 2023}, 2023, pp. 2433--2437.

\bibitem{vaswani2017attention}
A.~Vaswani, ``Attention is all you need,'' \emph{Advances in Neural Information Processing Systems}, 2017.

\bibitem{MCFN}
X.~Zhang and Y.~Li, ``A dual attention-based modality-collaborative fusion network for emotion recognition.''\hskip 1em plus 0.5em minus 0.4em\relax INTERSPEECH, 2023.

\bibitem{MSMSER}
S.~Wang, Y.~Ma, and Y.~Ding, ``Exploring complementary features in multi-modal speech emotion recognition,'' in \emph{ICASSP 2023-2023 IEEE International Conference on Acoustics, Speech and Signal Processing (ICASSP)}.\hskip 1em plus 0.5em minus 0.4em\relax IEEE, 2023, pp. 1--5.

\bibitem{zhao2023knowledge}
Z.~Zhao, Y.~Wang, and Y.~Wang, ``Knowledge-aware bayesian co-attention for multimodal emotion recognition,'' in \emph{ICASSP 2023-2023 IEEE International Conference on Acoustics, Speech and Signal Processing (ICASSP)}.\hskip 1em plus 0.5em minus 0.4em\relax IEEE, 2023, pp. 1--5.

\bibitem{ghosh2022mmer}
S.~Ghosh, U.~Tyagi, S.~Ramaneswaran, H.~Srivastava, and D.~Manocha, ``Mmer: Multimodal multi-task learning for speech emotion recognition,'' \emph{arXiv preprint arXiv:2203.16794}, 2022.

\bibitem{ma2023transformer}
H.~Ma, J.~Wang, H.~Lin, B.~Zhang, Y.~Zhang, and B.~Xu, ``A transformer-based model with self-distillation for multimodal emotion recognition in conversations,'' \emph{IEEE Transactions on Multimedia}, 2023.

\bibitem{shou2024low}
Y.~Shou, H.~Liu, X.~Cao, D.~Meng, and B.~Dong, ``A low-rank matching attention based cross-modal feature fusion method for conversational emotion recognition,'' \emph{IEEE Transactions on Affective Computing}, 2024.

\bibitem{pan2024gemo}
Y.~Pan, Y.~Hu, Y.~Yang, W.~Fei, J.~Yao, H.~Lu, L.~Ma, and J.~Zhao, ``Gemo-clap: Gender-attribute-enhanced contrastive language-audio pretraining for accurate speech emotion recognition,'' in \emph{ICASSP 2024-2024 IEEE International Conference on Acoustics, Speech and Signal Processing (ICASSP)}.\hskip 1em plus 0.5em minus 0.4em\relax IEEE, 2024, pp. 10\,021--10\,025.

\bibitem{kim2023focus}
K.~Kim and N.~Cho, ``Focus-attention-enhanced crossmodal transformer with metric learning for multimodal speech emotion recognition,'' in \emph{Proc. Interspeech 2023}, 2023, pp. 2673--2677.

\bibitem{liu2019roberta}
Y.~Liu, ``Roberta: A robustly optimized bert pretraining approach,'' \emph{arXiv preprint arXiv:1907.11692}, vol. 364, 2019.

\bibitem{MMRBN}
X.~Chen, ``Mmrbn: Rule-based network for multimodal emotion recognition,'' in \emph{ICASSP 2024-2024 IEEE International Conference on Acoustics, Speech and Signal Processing (ICASSP)}.\hskip 1em plus 0.5em minus 0.4em\relax IEEE, 2024, pp. 8200--8204.

\bibitem{zou2022speech}
H.~Zou, Y.~Si, C.~Chen, D.~Rajan, and E.~S. Chng, ``Speech emotion recognition with co-attention based multi-level acoustic information,'' in \emph{ICASSP 2022-2022 IEEE International Conference on Acoustics, Speech and Signal Processing (ICASSP)}.\hskip 1em plus 0.5em minus 0.4em\relax IEEE, 2022, pp. 7367--7371.

\bibitem{chen2022wavlm}
S.~Chen, C.~Wang, Z.~Chen, Y.~Wu, S.~Liu, Z.~Chen, J.~Li, N.~Kanda, T.~Yoshioka, X.~Xiao \emph{et~al.}, ``Wavlm: Large-scale self-supervised pre-training for full stack speech processing,'' \emph{IEEE Journal of Selected Topics in Signal Processing}, vol.~16, no.~6, pp. 1505--1518, 2022.

\bibitem{MFDR}
Z.~Zhao, T.~Gao, H.~Wang, and B.~Schuller, ``Mfdr: Multiple-stage fusion and dynamically refined network for multimodal emotion recognition,'' in \emph{Proc. Interspeech 2024}, 2024, pp. 3719--3723.

\bibitem{jiang2023empathy}
L.~Jiang, D.~Wu, B.~Mao, Y.~Li, and W.~Slamu, ``Empathy intent drives empathy detection,'' in \emph{Proceedings of the 2023 Conference on Empirical Methods in Natural Language Processing}, 2023, pp. 6279--6290.

\bibitem{aljundi2017expert}
R.~Aljundi, P.~Chakravarty, and T.~Tuytelaars, ``Expert gate: Lifelong learning with a network of experts,'' in \emph{Proceedings of the IEEE conference on computer vision and pattern recognition}, 2017, pp. 3366--3375.

\bibitem{huang2023pram}
Y.~Huang, W.~Liu, X.~Zhang, J.~Lang, T.~Gong, and C.~Li, ``Pram: An end-to-end prototype-based representation alignment model for zero-resource cross-lingual named entity recognition,'' in \emph{Findings of the Association for Computational Linguistics: ACL 2023}, 2023, pp. 3220--3233.

\bibitem{busso2008iemocap}
C.~Busso, M.~Bulut, C.-C. Lee, A.~Kazemzadeh, E.~Mower, S.~Kim, J.~N. Chang, S.~Lee, and S.~S. Narayanan, ``Iemocap: Interactive emotional dyadic motion capture database,'' \emph{Language resources and evaluation}, vol.~42, pp. 335--359, 2008.

\bibitem{poria2018meld}
S.~Poria, D.~Hazarika, N.~Majumder, G.~Naik, E.~Cambria, and R.~Mihalcea, ``Meld: A multimodal multi-party dataset for emotion recognition in conversations,'' \emph{arXiv preprint arXiv:1810.02508}, 2018.

\bibitem{MER-HAN}
S.~Zhang, Y.~Yang, C.~Chen, R.~Liu, X.~Tao, W.~Guo, Y.~Xu, and X.~Zhao, ``Multimodal emotion recognition based on audio and text by using hybrid attention networks,'' \emph{Biomedical Signal Processing and Control}, vol.~85, p. 105052, 2023.

\bibitem{DWFORMER}
S.~Chen, X.~Xing, W.~Zhang, W.~Chen, and X.~Xu, ``Dwformer: Dynamic window transformer for speech emotion recognition,'' in \emph{ICASSP 2023-2023 IEEE International Conference on Acoustics, Speech and Signal Processing (ICASSP)}.\hskip 1em plus 0.5em minus 0.4em\relax IEEE, 2023, pp. 1--5.

\bibitem{LSTM-Attn}
S.~Mirsamadi, E.~Barsoum, and C.~Zhang, ``Automatic speech emotion recognition using recurrent neural networks with local attention,'' in \emph{2017 IEEE International conference on acoustics, speech and signal processing (ICASSP)}.\hskip 1em plus 0.5em minus 0.4em\relax IEEE, 2017, pp. 2227--2231.

\bibitem{KS-Transformer}
W.~Chen, X.~Xing, X.~Xu, J.~Yang, and J.~Pang, ``Key-sparse transformer for multimodal speech emotion recognition,'' in \emph{ICASSP 2022-2022 IEEE International Conference on Acoustics, Speech and Signal Processing (ICASSP)}.\hskip 1em plus 0.5em minus 0.4em\relax IEEE, 2022, pp. 6897--6901.

\bibitem{DST}
W.~Chen, X.~Xing, X.~Xu, J.~Pang, and L.~Du, ``Dst: Deformable speech transformer for emotion recognition,'' in \emph{ICASSP 2023-2023 IEEE International Conference on Acoustics, Speech and Signal Processing (ICASSP)}.\hskip 1em plus 0.5em minus 0.4em\relax IEEE, 2023, pp. 1--5.

\bibitem{MSTR}
Z.~Li, X.~Xing, Y.~Fang, W.~Zhang, H.~Fan, and X.~Xu, ``Multi-scale temporal transformer for speech emotion recognition,'' \emph{arXiv preprint arXiv:2410.00390}, 2024.

\bibitem{MM-DFN}
D.~Hu, X.~Hou, L.~Wei, L.~Jiang, and Y.~Mo, ``Mm-dfn: Multimodal dynamic fusion network for emotion recognition in conversations,'' in \emph{ICASSP 2022-2022 IEEE International Conference on Acoustics, Speech and Signal Processing (ICASSP)}.\hskip 1em plus 0.5em minus 0.4em\relax IEEE, 2022, pp. 7037--7041.

\bibitem{xu2025gatem}
W.~Xu, Z.~Dong, R.~Wang, X.~Xu, and Z.~Zhang, ``Gatem 2 former: Gated feature selection and expert modeling in multimodal emotion recognition,'' in \emph{ICASSP 2025-2025 IEEE International Conference on Acoustics, Speech and Signal Processing (ICASSP)}.\hskip 1em plus 0.5em minus 0.4em\relax IEEE, 2025, pp. 1--5.

\bibitem{SAAVN}
Y.~Yu, W.~Huang, F.~Sun, C.~Chen, Y.~Wang, and X.~Liu, ``Sound adversarial audio-visual navigation,'' in \emph{The Tenth International Conference on Learning Representations, {ICLR} 2022, Virtual Event, April 25-29, 2022}.\hskip 1em plus 0.5em minus 0.4em\relax OpenReview.net, 2022.

\end{thebibliography}

\vspace{12pt}
\color{red}

\end{document}